\documentstyle[12pt,aasms4]{article} 
\def\etal{{et~al.}\ }

\def\msun{{\rm\,M_\odot}}

\def\vol#1  {{{#1}{\rm,}\ }}

\def\etal{et al.\ }

\def\clock{\count0=\time \divide\count0 by 60
     \count1=\count0 \multiply\count1 by -60 \advance\count1 by \time
     \number\count0:\ifnum\count1<10{0\number\count1}\else\number\count1\fi}

\begin{document}
\title{A Possible Lateral Gamma-Ray Burst Jet from Supernova 1987A}
\author{Renyue Cen\altaffilmark{1}}

\altaffiltext{1} {Princeton University Observatory, Princeton University, Princeton, NJ 08544; cen@astro.princeton.edu}

\begin{abstract}
There was a bright, transient companion spot to SN1987A
with a projected distance of about 
17 light-days, observed by optical speckle interferometry
one to two months after explosion.
It is shown here that the bright spot may be due to
a receding ultra-relativistic jet 
traveling at $\sim 53^\circ$ 
to the observer-to-SN1987A vector, through
a circumstellar medium of density profile $\rho (r)\propto r^{-2}$.
If it had approached us along the line of sight,
a very bright gamma-ray burst would have been seen
with an apparent isotropic energy of $\sim 10^{54}~$erg
and an openning angle of a few degrees.
The model provides an adequate explanation
for the evolution of the spot, although there are still
problems in explaining its observed color.
This model implies that at least some GRBs would 
be seen as going through 
a medium with density $\rho(r)\propto r^{-2}$
rather than a uniform medium,
which is frequently adopted in GRB calculations.
Improved analysis of the speckle data
has revealed another and fainter spot on the opposite side.
\end{abstract}


\section{Introduction}
The SN1987A in the Large Magellanic Cloud 
was a rare and unique event thanks to 
its nearness to us. 
It has been observed with all 
available modern instruments since its explosion (e.g.,
Chevalier 1992)
and is expected to have another magnificent display in a few years
when the expanding ejecta hits the circumstellar ring 
(e.g., Borkowski, Blondin, \& McCray 1997).
Perhaps one of the greatest mysteries about SN1987A
is the mysterious bright companion spot 
that was observed
by optical speckle interferometry
(Nisenson \etal 1987, N87 hereafter;
Meikle, Matcher, \& Morgan 1987, M87 hereafter)
about one month after the SN1987A explosion,
with a projected displacement from SN1987A of about 17 light days.
Its close proximity to SN1987A,
the fact that it was seen for only a few weeks,
and its high 
brightness (about one-tenth of the brightness of SN1987A itself)
make it certain that the spot was related to SN1987A itself.
Several models were proposed soon after its discovery
(Burrow \& Subramanian 1987;
Rees 1987;
Piran \& Nakamura 1987;
Goldman 1987;
Felten, Dwek, \& Viegas-Aldrovandi 1989)
but close examination showed that
there are formidable difficulties 
with all these models (Phinney 1988).

Recently, there was an interesting development in the observations
of gamma-ray bursts:
the supernova 1998bw was observed (Kulkarni \etal 1998b)
to coincide spatially
and temporally with the gamma-ray burst GRB980425.
This has led to suggestions
that gamma-ray bursts (GRBs) 
and supernovae (SNe) may be related
(Wang \& Wheeler 1998; Cen 1998).
Energetics dictate that if SNe are 
responsible for producing GRBs, 
GRBs have to be beamed, that is, GRBs
are jets from SNe.
Independently but consistently,
it is also required that
the jets have a beaming angle of a few degrees
in order to reconcile the high
rate of SN events with the low rate of GRB events.
The pressing question that arises then
is how to test this scenario,
where the vast majority of SN jets
would travel laterally and would not be seen as GRBs
due to the small beaming angle.
It is the goal of this {\it Letter} to examine the properties
of such lateral jets, suggesting 
that the observed bright companion spot of SN1987A
may be caused by such a jet from SN1987A.

\section{A Possible GRB Jet from SN1987A}
The bright SN1987A companion spot was observed
independently by two groups
(N87; M87).
It was observed 
at H$_\alpha$ and several other
optical wavelengths using speckle interferometry
by the CfA group (N87)
on days 30 and 38 after the SN1987A explosion 
at a separation of $0^{''}.059\pm 0^{''}.008$ from SN1987A.
Adopting a fiducial value of $50$kpc 
for the distance to SN1987A
(Panagia \etal 1991; 
Gould 1995; 
Sonneborn \etal 1997;
Lundqvist 1999),
one obtains a perpendicular
separation of $r_\perp=17~$light-days.
Assuming that the spot
was due to an ultra-relativistic jet leaving SN1987A
at the time of the explosion,
it gives a travel time $\Delta t=34~$days
and yields an apparent perpendicular
velocity of $v_\perp=0.5c$ ($c$ is the speed of light).
Because $v_\perp=c \sin\theta/(1+\cos\theta)$,
where $\theta$ is the angle between the jet direction and
the observer-SN1987A vector,
one finds $\theta=53^\circ$.
Thus, if the spot was due to the working surface 
of a relativistic jet, the jet was a receding one!
The spot detected by M87
on day 50 at a separation
$0^{''}.074\pm 0^{''}.008$ is fully consistent
with the observations of N87
for a jet traveling at near the speed of light.
Interestingly, new image reconstructions from
the CfA speckle data show possible indications of a second, weaker
jet, with a larger separation, on the opposite side of the SN1987A
(Nisenson \& Papaliolios 1999). 
Although working surface models were disfavored earlier (Phinney 1988),
in light of this new observation of a counter jet
and possible association of supernovae with GRBs (see \S 1),
it seems worthwhile to re-examine this type of models
in the context of GRB jets.

Let us now examine the spectral properties 
of an ultra-relativistic GRB jet (Cen 1998).
The jet can be characterized by its
initial equivalent isotropic energy $E_{iso}$, initial coasting Lorentz
factor $\Gamma_i$ and opening solid angle $\Omega$.
For the current analysis only
an external shock model (Rees \& M\'esz\'aros 1992)
is considered for the jet. 
The reverse shock is not considered).
It is assumed that the external shocked electrons
have a power-law distribution function:
\begin{equation}
N(\Gamma_e)d\Gamma_e=A(t)\Gamma_e^{-p}d\Gamma_e, 
\end{equation}
\noindent where $\Gamma_e$ is the Lorentz factor
of electrons in the jet comoving frame,
and $A(t)$ is a coefficient (to be determined)
that is assumed 
to be a function of time only.
Time $t$ measured in the burster frame
is used as the time variable to express various quantities
in the derivations, but the final results
are converted to be shown using observer's time.
We will only consider synchrotron radiation from the shock heated electrons.
For the analysis below we will assume that $p>1$
(Tavani 1996) so
the integral of equation (1) is convergent at the high end.
We set
\begin{equation}
\int_{\Gamma_{e}}^\infty N(\Gamma_e^\prime)d\Gamma_e^\prime=\Omega r^2 c n t_{cool}\Gamma(r),
\end{equation}
\noindent 
where $n$ the number density of the external medium into which
the shock is propagating,
$r$ is the distance of the shock from SN1987A 
($r$ and $t$ are used interchangeably throughout the paper assuming $r=ct$)
and
$t_{cool}$ is the electron cooling time (see equation [4]).
Equation (2) is equivalent to stating that 
the number of electrons with $\Gamma>\Gamma_{e}$
at time $t$
is the number of electrons that have been shocked 
within the last $t_{cool}$ time interval,
and earlier shocked electrons have cooled to lower energies.
The last factor $\Gamma(r)$ on the right hand side
of equation (2) accounts for the time boost of a moving object.
Integrating equation (2) yields
\begin{equation}
A(t) = (p-1)\Omega r^2 c n t_{cool} \Gamma_{e}^{p-1}(r) \Gamma(r)\ .
\end{equation}
\noindent 
The synchrotron cooling time measured
in the comoving frame for an electron with $\Gamma_{e}$ is
\begin{equation}
t_{cool}={\Gamma_{e} m_e c^2\over P_{e}},
\end{equation}
\noindent 
The majority of the freshly shocked electrons (as we will adopt $p\sim 6$)
have a Lorentz factor 
\begin{equation}
\Gamma_{e}(r)=\Gamma(r){m_p\over m_e}\xi_e,
\end{equation}
\noindent 
where $\Gamma(r)$ is the shock Lorentz factor,
$m_p$ and $m_e$ are proton and electron mass
and $\xi_e$ is an equipartition parameter (Waxman 1997).
The synchrotron radiation power, $P_{e}$,  
for an average electron with $\Gamma_{e}$ in a randomly directed
magnetic field $B$ is (Blumenthal \& Gould 1970):
\begin{equation}
P_{e}={4\over 3} \sigma_T c \Gamma_{e}^2 {B^2\over 8\pi},
\end{equation}
\noindent 
where $\sigma_T=6.6\times 10^{-25}$cm$^2$ is the Thomson cross section.
$B$ (Waxman 1997) is linked to
the energy density of the postshock external nucleons,
$4\Gamma(r)^2 n m_p c^2$, by
\begin{equation}
{B^2\over 8\pi} = 4\Gamma(r)^2 n m_p c^2 \xi_B,
\end{equation}
\noindent 
where $\xi_B$ is the equipartition parameter for the magnetic field.

Now we may proceed to obtain the total emission. 
For the present purpose
it is adequate to assume that
the spectral emissivity of each electron is a delta
function $P_\nu=P_e \delta (\nu-\nu_e)$, where
$P_e$ can be expressed by equation (6), and the characteristic 
synchrotron radiation frequency $\nu_e$ 
for electrons with $\Gamma_e$ is (Rybicki \& Lightman 1979) 
\begin{equation}
\nu_e = \Gamma_e^2 {e B\over 2\pi m_e c}\ .
\end{equation}
\noindent 
Multiplying equation (1) by $P_\nu$ and integrating over $\Gamma_e$,
and using equations (3,4,6,7,8)
give the total emission in the comoving frame
\begin{equation}
j(\nu,t) = {1\over 2} (p-1) \Omega r^2 n(r) m_e c^3 \Gamma_e(r)\Gamma(r)\nu_{e}^{-1}(r) \left({\nu\over\nu_e}\right)^{-{p-1\over 2}}\ .
\end{equation}
\noindent 
It is noted that the above expression for $j(\nu,t)$ is
valid only above a lower cutoff frequency, $\nu_l$, 
since the total energy has to be finite.
We observe the following simple
ansatz to obtain $\nu_l$:
the total radiation emitted during the time interval $t_{cool}$ (in 
the comoving frame) should not exceed the total energy input 
to the thermalized electrons during 
the same time interval, 
which translates to the following relation:
\begin{equation}
\int_{\Gamma_{e}}^\infty \Gamma_e^\prime m_e c^2 N(\Gamma_e^\prime)d\Gamma_e^\prime= t_{cool} \int_{\nu_{l}}^\infty j(\nu,t) d\nu.
\end{equation}
\noindent 
Integrating both sides of equation (10) and using equations (3,4,6,7,8)
yield
\begin{equation}
\nu_l(t)=\left({p-2\over p-3}\right)^{2\over p-3} \nu_{e}(t),
\end{equation}
\noindent 
where $\nu_e$ is given by equation (8).
Note that the derived $\nu_l(t)$ is slightly larger than $\nu_e(t)$.
Below $\nu_{l}$, $j(\nu,t)$ scales as 
\begin{equation}
j(\nu,t)=j(\nu_{l},t) ({\nu\over \nu_{l}})^{1/3}.
\end{equation}
\noindent 
Synchrotron self-absorption becomes important
only at lower frequencies than those of interest here
and is thus ignored in the present analysis.


In order to compute $j(\nu,t)$ as a function of time,
one needs to specify the circumstellar
medium density distribution
and the evolution of the bulk Lorentz
factor of the shock.
The standard steady wind model
for the distribution of the circumstellar medium
of a red supergiant
is adopted:
\begin{equation}
\rho(r) = {\dot M\over 4\pi v_w r^2},
\end{equation}
\noindent 
where $\dot M$ is the mass loss rate of the star
and $v_w$ is the wind velocity.
Using $\dot M=4\times 10^{-5}\msun$yr$^{-1}$ and 
$v_w=10$km/s, as inferred from analysis of SN 1993J
(Fransson, Lundqvist, \& Chevalier 1996) yields
\begin{equation}
n(r) = \left({r\over r_0}\right)^{-2} \hbox{atoms/cm$^3$}
\end{equation}
\noindent 
with $r_0=1.1\times 10^{19}\hbox{cm}$.
This adopted density distribution is in fact quite consistent
with the measured circumstellar density 
of SN1987A (e.g., Sonneborn \etal 1998).
It is assumed that radiative losses are small,
which is appropriate at the later times of the fireball evolution
of interest here.
Then, for our adopted
$\rho(r)$, we find the following
scaling solution for $\Gamma(t)$ (Blandford \& McKee 1976)
\begin{equation}
\Gamma(t)=\Gamma_i(t/t_{dec})^{-1/2}
\end{equation}
\noindent 
for $t>t_{dec}$. 
For $t\le t_{dec}$ we simply set
$\Gamma(t)=\Gamma_i$.
The transition time $t_{dec}$, measured in the burster frame,
is set to be that when
the mass of the swept-up circumstellar medium
is equal to $1/\Gamma_i$ of the initial fireball rest mass,
yielding
\begin{equation}
t_{dec}={E_{iso}\over 4\pi m_p \Gamma_i^2 c^3 r_0^2}.
\end{equation}
\noindent 

The flux density (in units of erg~cm$^{-2}$~s$^{-1}$~Hz$^{-1}$)
of the jet at the observer at observed frequency $\nu_{obs}$ 
at observer's time $t_{obs}$ is
(Blandford \& Konigl 1979)
\begin{equation}
S_\nu (\nu_{obs},t_{obs}) ={1\over 4\pi d_{SN}^2} j\left({\nu_{obs}\over D},{t_{obs}\over 1+\cos \theta}\right) D^3\left({t_{obs}\over 1+\cos\theta}\right),
\end{equation}
\noindent 
where $D(t)\equiv (1+\beta\cos\theta)^{-1}\Gamma^{-1}(t)$
is the Doppler factor of the moving surface.
Flux density is then converted to magnitude
to compare with observations.
Figure 1 shows the magnitudes 
of the jet at $6560\AA$ (solid curve) and $4500\AA$ (dashed curve),
as a function of time measured in the observer's frame, $t_{obs}$
[note $t_{obs}=t (1+\cos\theta)$].
Note that the open circle at day 98 is from a
recent re-analysis of the observational data (Nisenson 1999). 
The observed points have been dereddened for extinction using 
the observed color excess $E(B-V)=0.19$ for SN1987A
(Fitzpatrick \& Walborn 1990)
and the extinction curve given by
Seaton (1979).
The following parameter values are used
for the results shown in Figure 1:
%
$\xi_e=1/3$, $\xi_B=1/4$,
$p=6.0$, $E_{iso}=2\times 10^{54}$erg,
$\Gamma_i=300$,
$\Omega=1.5\times 10^{-3}~$sr, $\theta=53^\circ$ and $d_{SN}=50~$kpc.
All the parameters used are characteristic 
of a supernova GRB jet proposed (Cen 1998)
and are consistent with known GRB observations.
Note that $E_{iso}=10^{54}$erg is
capable of accounting
for the most luminous GRBs observed
(e.g., GRB971214, Kulkarni \etal 1998a).
A detailed analysis of the jet in the context of
a GRB and its afterglows will be given elsewhere.

The GRB jet model fits the speckle observations of 
the spot at $6560\AA$ reasonably well over the entire
period where observational data are available.
However, the model appears to be 
too ``blue" in the sense,
i.e., it appears to be too bright at shorter wavelengths.
For example, the computed spot at $4500\AA$ appears
to be too bright by about two magnitudes compared to
the observed spot.
While the model is consistent (not shown in the figure)
with infared observations
of SN 1987A (e.g., at $4.6\mu$m, Bouchet \etal 1987),
it also appears to be too bright in the UV 
compared to the total flux of SN 1987A (e.g., $3100\AA$, Kirshner 1987)
by about a factor of ten, consistent with Phinney (1988).
Clearly, more work is needed to improve upon this simple model.
One way to avoid excess flux at short wavelengths
is to introduce a large, intrinsic color excess,
say, $E(B-V)\sim 1.5$.

The sharp turn near days 30-40 is due to
the sharp turn in the spectrum at $\nu_l(t)$.
The peak of the evolution of the jet brightness 
at a given wavelength corresponds to the epoch
when $\nu(t) = \nu_l(t)$
and the sharp turn (to faint) of brightness of the jet
at earlier times is primarily due to the fact that
the $D^3$ term in equation (17) goes roughly
as $t^{3/2}$ 
and $\nu_l$ increases rapidly with decreasing time (roughly $\propto t^{5/2}$)
combined with the spectral form of $\nu^{1/3}$ below $\nu_l$.
The evolution of the brightness of the jet past
the peak is primarily determined by the combined effect
of the evolution of $\nu_{l}$ and $p$.
The quantity $p$ is well constrained by the observed evolution
of the optical spot.
We find that $p\sim 6$ is required in order to
provide an acceptable fit to the observed
optical spot.
A larger $p$ ($>7$) would produce too steep a decline
around $t_{obs}\sim 30~$days.
A smaller $p$ ($<5$) would produce a flat to rising
temporal evolution and is inconsistent with 
the observation, i.e., the spot should have been visible longer.

This ``counterspot" on the 
opposite side (Nisenson \& Papaliolios 1999)
has an apparent separation of $0^{''}.16$
at the same time when the first spot was seen,
giving an apparent {\it superluminal}
perpendicular velocity of $v_{\perp, second}=1.36c$.
If one assumes that this weaker jet 
was in the exact opposite direction from the first
(i.e., $\theta_{second}=180-\theta=127^\circ$),
it is required that $v_{second}=0.84c$.
However, due to the uncertainties in $d_{SN}$,
it is possible that $v\sim c$ may be allowed for both jets.
It is interesting and should be emphasized 
that the two jets have unequal strengths,
a prediction of the model proposed by Cen (1998) to account for
the asymmetrical natal kick of neutron stars (pulsars).
The asymmetrical pair of jets 
would induce star-recoil 
 with the induced bulk velocity of
the star being $650 (m_{star}/10M_\odot)^{-1}~$km/s (Cen 1998),
which moves about $0^".03(m_{star}/10M_\odot)^{-1}$ in ten years
(using $d_{SN}=50$kpc).
This effect might be observable by detecting a shift of
the position of the neutron star/pulsar
or the centroid of the debris (Garnavich 1999).
Based on available debris data
(Haas \etal 1990; 
Spyromilio, Meikle, \& Allen 1990;
Jennings \etal 1993;
Utrobin, Chugai, \& Andronova 1995;
Wang \etal 1996),
if seems that the debris
does {\it not} share 
the recoil movement of the star
but shares the movement of the jet.


\section{Conclusion}
It is shown here that the bright companion spot of SN1987A
may be due to a receding ultra-relativistic jet 
traveling at $\sim 53^\circ$ to the observer-to-SN1987A vector,
through a circumstellar
medium with a stellar wind like density $\rho(r)\propto r^{-2}$.
The model provides an adequate explanation
for the evolution of the observed optical companion spot,
at least energetically,
although more modeling is required
to produce a satisfactory color of the spot.
The parameters for the jet are
characteristic of or required by 
the observed GRBs (with
$E_{iso}=2\times 10^{54}$erg,
$\Gamma_i=300$) with an openning angle of a few degrees.
If the jet traveled towards us along the line of sight,
a very bright GRB would be seen
with an inferred isotropic energy of $\sim 10^{54}~$erg.
If this model is correct,
it implies that at least some GRBs would 
be seen as going through a medium with density $\rho(r)\propto r^{-2}$,
rather than a uniform density medium.
It is urgent to systematically search 
for GRB-supernova associations or supernova-jet associations
in order to test this hypothesis.

\acknowledgments
I thank Arlin Crotts, Dick McCray and 
Pete Nisenson for stimulating discussions,
Jeremy Goodman for suggesting looking for
evidence of lateral GRB jets and
Micheal Strauss for carefully reading the manuscript.
I want to thank the second referee, Jim Felten,
for working tirelessly to help improve the paper.
The work is supported in part
by grants NAG5-2759 and AST93-18185, ASC97-40300.

{}

\clearpage
\figcaption[FLENAME]{
shows the magnitudes 
of the jet at $6560\AA$ (solid curve) and $4500\AA$ (dashed curve),
as a function of time measured in the observer's frame, $t_{obs}$.
The origin of the $t_{obs}$ coincides with the time of
the SN1987A explosion.
The symbols are the observed magnitudes (dereddened)
of bright companion spot at 
$6560\AA$ at days 30 and 38
(filled circles; N87), at $4500\AA$ at day 38 (filled square; N87)
at $6585\AA$ at day 50 (filled triangle; M87)
and at $6560\AA$ at day 98 (open circle; N99).
\label{fig1}}

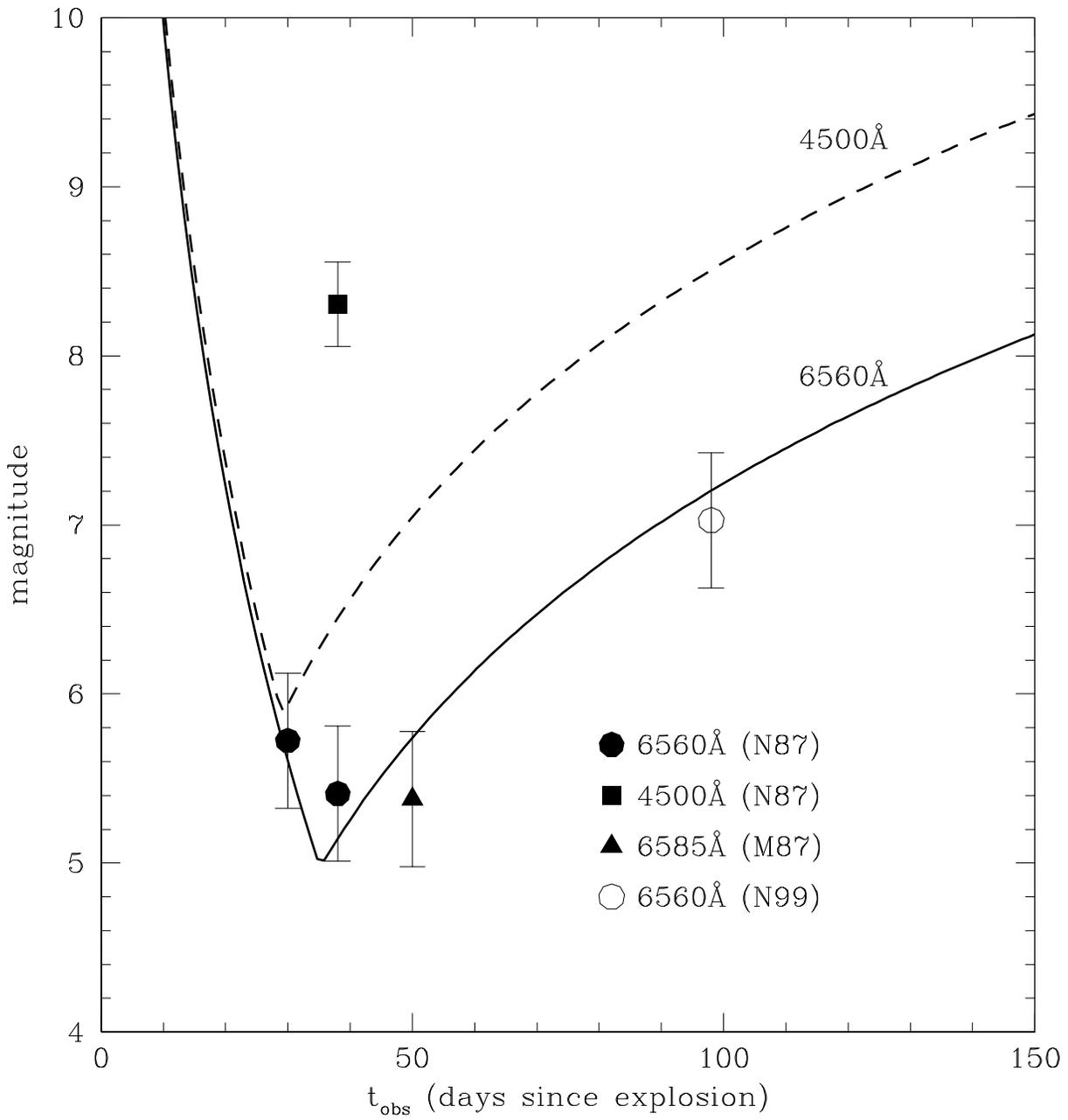
\end{document}